\documentclass[pre,aps,twocolumn]{revtex4}

\usepackage{graphicx}
\usepackage{amsmath}
\usepackage{amssymb}
\usepackage{ifthen}
\usepackage{booktabs}
\usepackage{color}

\usepackage{graphicx}
\usepackage{dcolumn}
\usepackage{bm}
\usepackage{longtable}

\newcommand{\bb}{\begin{equation}}
\newcommand{\ee}{\end{equation}}
\newcommand{\ba}{\begin{eqnarray*}}
\newcommand{\ea}{\end{eqnarray*}}
\newcommand{\rhor}{\rho({\bf r})}
\newcommand{\dd}{{\rm d}}
\newcommand{\rr}{{\mathbf r}}
\newcommand{\dr}{{\rm d}{\bf r}}

\begin{document}

\title{The Influence of Intermolecular Forces at Critical Point Wedge Filling}

\author{Alexandr \surname{Malijevsk\'y}}
\affiliation{{Department of Physical Chemistry, Institute of Chemical Technology, Prague, 166 28 Praha 6, Czech Republic;}\\
 {Institute of Chemical Process Fundamentals, Academy of Sciences, 16502 Prague 6, Czech Republic}}
\author{Andrew O. \surname{Parry}}
\affiliation{Department of Mathematics, Imperial College London, London SW7 2B7, UK}

\begin{abstract}
We use microscopic density functional theory to study filling transitions in systems with long-ranged wall-fluid and short-ranged fluid-fluid  forces occurring in a
right-angle wedge. By changing the strength of the wall-fluid interaction we can induce both wetting and filling transitions over a wide range of temperatures and study
the order of these transitions. At low temperatures we find that both wetting and filling transitions are first-order in keeping with predictions of simple local
effective Hamiltonian models. However close to the bulk critical point the filling transition is observed to be continuous even though the wetting transition remains
first-order and the wetting binding potential still exhibits a small activation barrier. The critical singularities for adsorption for the continuous filling transitions
depend on whether retarded or non-retarded wall-fluid forces are present and are in excellent agreement with predictions of effective Hamiltonian theory even though the
change in the order of the transition was not anticipated.
\end{abstract}

\maketitle

The properties of confined and inhomogeneous fluids have received enormous theoretical and experimental attention over the last few decades
\cite{evans_79,evans90,sullivan,dietrich,schick,binderrev,bonn,saam}. These have revealed convincing evidence that the interplay between surface tension, intermolecular
forces and the substrate geometry can induce new phase transitions and fluctuation regimes beyond that occurring for bulk fluids. A simple but striking example of a
transition induced by a structured surface is the filling transition pertinent to fluid adsorption in a linear wedge formed by the junction of two planar walls with
opening angle $2\psi$. Simple macroscopic arguments dictate that, at bulk two phase coexistence, a wedge-gas interface is completely filled by liquid for temperatures
$T>T_{\rm fill}$, where the filling temperature $T_{\rm fill}$ is given implicitly by
 \cite{shuttle, concus, pomeau, hauge}
\begin{equation}
\theta(T_{\rm fill})=\frac{\pi}{2}-\psi \label{thermo}
\end{equation}
where $\theta(T)$ denotes the equilibrium contact angle. Since (\ref{thermo}) is exact, all fluids that form drops  with a finite contact angle will exhibit a filling
transition in a wedge tuned to the appropriate opening angle. The properties of the filling transition were first studied using mesoscopic effective Hamiltonian models
which concluded that the transition may be first-order or continuous determined by the qualitative properties of the interfacial binding potential (see below)
\cite{rejmer,wood1,wood2}. The critical singularities characteristic of continuous filling depend sensitively on the dimensionality and the range of the intermolecular
forces. In 3D, even in the presence of long-ranged forces, they show dramatically enhanced interfacial fluctuation effects compared to wetting
\cite{wood2,binder03,binder05} while in 2D many properties of wetting and filling are precisely related by a symmetry called wedge covariance \cite{Cov}. Interestingly
in 3D the conditions for continuous wedge filling appear to be less restrictive than those for continuous wetting hinting that they should be more easily observed in the
laboratory \cite{mistura}. Several of the early predictions of the effective Hamiltonian theory of filling for systems with short-ranged forces have been confirmed in
more microscopic approaches including simulations \cite{binder03,binder05}, exactly solvable models \cite{parry05,abraham02,abraham03} and field theory \cite{delfino}.
However recent studies using microscopic Fundamental Measure Density Functional Theory (FM-DFT) suggest that with long-ranged forces there are new aspects of 3D wedge
filling not anticipated by interfacial models \cite{MP1,MP2}. In particular there is evidence that in acute wedges or close to the bulk critical temperature $T_c$
continuous filling transitions are even more prevalent than first thought. This phenomenon of critical point wedge filling is not easily explicable using interfacial
models, which are typically limited to open wedges and to low temperatures and neglect completely short-ranged correlations associated with the liquid structure near the
walls. In this paper we show that critical point wedge filling is present for a broader class of wall-fluid intermolecular forces and determine numerically the critical
singularity characterising the divergence of the adsorption which is found to be in excellent agreement with the predicted value for both retarded and non-retarded van
der Waals forces.

\begin{figure}[ht]
\includegraphics[width=8.5cm]{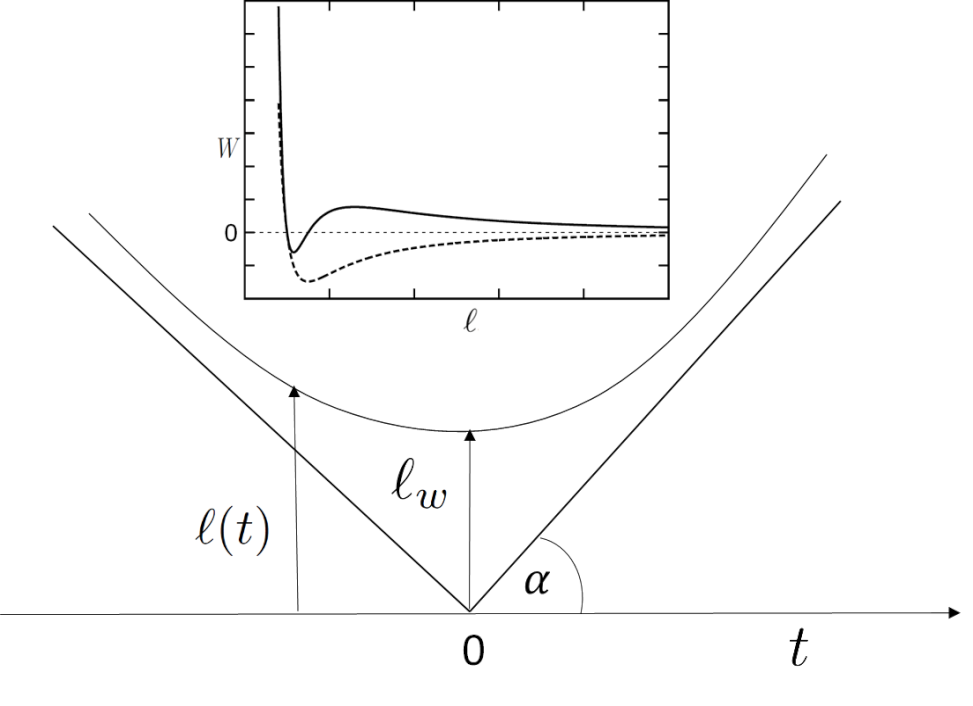}
\caption{Schematic illustration of the meniscus height $\ell(t)$ in a linear wedge with a tilt angle $\alpha$ with $\ell_w$ the equilibrium mid-point
height measured above the apex. Inset shows the qualitative shape of the binding potential $W(\ell)$
 for wetting at a planar wall for the cases of first-order wetting (solid line) and continuous (dashed line) wetting transitions.}\label{fig1}
\end{figure}

To begin we recall the simple effective Hamiltonian theory of wedge filling valid for shallow wedges with small tilt angle $\alpha=\pi/2-\psi$
\cite{rejmer,wood2}. Assuming translational invariance along the wedge, the position of the wall may be approximated $z_{w}=\alpha|t|$ where $t$
denotes the horizontal coordinate (see Fig.~\ref{fig1}) and where we used a standard shallow-wedge approximation $\tan\alpha\approx\alpha$. If
$\ell(t)$ denotes the local height of the liquid-gas interface above the horizontal then the free-energy cost of a given interfacial configuration is
described by the approximate functional
\begin{equation}
F[\ell]=\int dt \left\{\frac{\gamma}{2} \dot\ell^2+W(\ell-z_{w})\right\}.
\label{Heff}
\end{equation}
Here $\gamma$ is the liquid-gas surface tension, $\dot\ell\equiv \frac{d\ell}{dt}$ and $W(\ell)$  is the interfacial binding potential modelling the
local wetting properties of a planar wall-gas interface. At mean-field level, simple minimization of $F[\ell]$ determines the equilibrium profile and
that the height $\ell_w$ of the interface above the wedge apex satisfies
\begin{equation}
\frac{\gamma}{2}(\alpha^2-\theta^2)=W(\ell_w).
\label{EL}
\end{equation}
Now suppose we are at bulk two-phase coexistence (chemical potential $\mu=\mu_{\rm sat}(T)$) on the vapor side and below the wetting temperature
$T_{\rm wet}$ so that $\theta>0$ implying $W(\ell)$ has a global minimum (at $\ell_\pi$, say) and decays to zero as $\ell\to\infty$
\cite{sullivan,dietrich,schick}. From (\ref{EL}) it follows that if $W(\ell)$ has an activation barrier (see inset Fig.~1) then the filling
transition occurring as $\theta\to\alpha^+$ is first-order so that $\ell_w$ jumps discontinuously from a finite to macroscopic value at $T_{\rm
fill}$. If there is no activation barrier then the filling transition is second-order and $\ell_w$ diverges continuously. This condition translates
as follows: Second-order wetting implies second-order (continuous) filling while first-order wetting implies first-order filling unless $T_{\rm
fill}<T_s$ where $T_s$ is the spinodal temperature,  defined as the lowest temperature at which the barrier in $W(\ell)$ appears. In this case the
filling transition will also be continuous. The present theory also predicts the critical exponents for continuous filling. For long-ranged forces,
the binding potential decays as $W(\ell)\sim A \ell^{-p}$ with $A$ negative for continuous filling and $p=2,3$ for non-retarded and retarded
dispersion forces respectively. This implies that the excess adsorption of liquid near the apex $\Gamma=\int\int \dd x \dd z
(\rho(x,z)-\rho_g)\propto \ell_w^2$ where $\rho_g$ is the bulk gas density, diverges as
\begin{equation}
\Gamma\sim (\theta-\alpha)^{-\frac{2}{p}}.
\label{Gamma}
\end{equation}
The same result for the divergence of $\Gamma$ follows from a different argument which determines directly the free-energy  cost of filling the wedge
with liquid to thickness $\ell_w$. Young's equation implies that the surface tensions contribute a term $\propto\sin(\theta-\alpha)\ell_w$ which is
balanced by a term $\propto \ell_w^{1-p}$ coming directly from the intermolecular forces  \cite{wood2}. If the latter contribution is net repulsive,
as it must be for continuous filling, it follows that the film thickness diverges continuously as $\ell_w\sim (\theta-\alpha)^{-\frac{1}{p}}$. These
mean-field arguments are believed to be valid for $p<4$ \cite{wood2}. For $p\ge4$ mean-field  theory breaks down since interfacial fluctuations are
dominant and the adsorption diverges with a universal power-law $\Gamma_{w}\sim (\theta-\alpha)^{-\frac{1}{2}}$ characteristic of filling in systems
with short-ranged forces. This latter prediction is consistent with extensive simulations of filling in the 3D Ising model \cite{binder03}.

An interesting theoretical aspect of wedge filling is that in attempting to understand the transition for more acute wedges we inevitably push
effective Hamiltonian theories to the limits of their applicability. It is of course straightforward to improve on the shallow wedge model
(\ref{Heff}) by replacing the square gradient term with a full ``drumhead" expression $\int dt\sqrt{1+\dot\ell^2}$ \cite{rejmer}. However to reliably
model filling in acute wedges also requires that we abandon the use of the planar wetting potential $W(\ell)$ and adopt a fully {\it{non-local}}
description of the interaction of the interface with the non-planar wall \cite{nolo1,nolo2,nolo3}. This would also account for the self-interaction
between the wetting films across the wedge  \cite{self}. While this is feasible at least numerically at low temperatures, by for example assuming a
sharp-kink approximation for the density profile \cite{dietrich,nolo3}, this is much more difficult if the filling transition occurs close to the
bulk critical temperature. To understanding filling in this regime, even at mean-field level, it is necessary and far simpler to move directly to
more microscopic DFT methods similar to studies of fluid adsorption in other geometries
\cite{tar_ev,evans86,stewart,nold,AM,AM2,petr,MP3,giac,singh}.

 Within the framework of classical DFT the equilibrium density profile is found by minimizing the grand potential functional \cite{evans_79}
\begin{equation}
\Omega[\rho]=F[\rho]+\int \dr\rhor[V(\rr)-\mu]
\end{equation}
where $F[\rho]$ is the intrinsic free energy
functional of the fluid one-body density, $\rho(\bf{r})$. Following a perturbative scheme modern DFT usually separates this as
\begin{equation}
F[\rho]=F_{\rm ideal}[\rho]+F_{\rm hs}[\rho]+\frac{1}{2}\int\int \dr_1 \dr_2\rho(\rr_1)\rho(\rr_2)u_{a}(r_{12})
\label{Fintrinsic}
\end{equation}
where the first two terms on the right hand side are the ideal and hard-sphere contributions respectively.  For the latter we use Rosenfeld's approximate FM theory which accurately models
short-ranged repulsive correlations between the fluid atoms \cite{ros,roth_fmt}. The final term in  (\ref{Fintrinsic}) is a mean-field treatment of the attractive part,
$u_{a}(r)$, of the intermolecular fluid-fluid potential.  Following our earlier study we take this to be a Lennard-Jones (LJ) potential $u_{a}(r)= -4\varepsilon
(\sigma/r)^6H(r-\sigma)$ which is truncated at $r_c=2.5\,\sigma$, where $\sigma$ is the hard-sphere diameter and $H(x)$ is the Heaviside function.

The external potential $V(\rr)$ arises from summing over all two-body wall-fluid interactions. That is
$V({\bf{r}})=\rho_w\int_\mathcal{V}d{\bf{r}}'\phi_{wf}(|{\bf{r}}-{\bf{r}}'|)$  where the integration  is over the whole domain $\mathcal{V}$ of the wall which is assumed
to be a uniform distribution of atoms, with density $\rho_w$. Here $\phi_{wf}(r)$ is the wall-fluid two-body interaction which for $r>\sigma$ we suppose is given by the
generalised long-ranged potential $\phi_w(r)=-4\varepsilon_w\left(\frac{\sigma}{r}\right)^{n}$. The exponent $n=4+p$ for purposes of comparison with the above effective
Hamiltonian theory. A hard-wall repulsion is also imposed for $r<\sigma$. For a {\it{planar wall}} occupying the half-space $z<0$, say, this integration produces the
external potential
\begin{equation}
V_\pi(z;n)=\frac{8\pi\varepsilon_w\rho_w\sigma^n}{(n-2)(3-n)}z^{3-n}\,;z>\sigma
\end{equation}
while for a right angle wedge ($\psi=\pi/4$) the potential $V({\bf{r}})=V(x,z)$ is a more complicated
function of Cartesians $x,z>0$ although it is translationally invariant along the wedge. Calculation shows that this can be written
$V(x,z;n)=V_\pi(z;n)+V_2(x,z;n)$ where the additional contribution
\begin{widetext}
 \begin{equation}
V_2(x,z;n)=C_n \left[\frac{I_{n-1}\left(\frac{z}{x}\right)+I_{n-1}(\infty)}{x^{n-3}}+\frac{I_{n-1}\left(\frac{x}{z}\right)-I_{n-1}(\infty)}{z^{n-3}}\right] \label{V2}
\end{equation}
\end{widetext}
 with $C_n=-\frac{4\sqrt{\pi}\varepsilon_w\rho_w\sigma^n\Gamma\left(\frac{n-1}{2}\right)}{\Gamma\left(\frac{n}{2}\right)(n-3)}$ and $I_n(x)\equiv\int
dx(1+x^2)^{-n/2}$. For integer $n$ the latter integral can be easily determined using the recurrence formula
\begin{equation}
 nJ_{n+2}(y)=\sinh y\cosh^{-n}y+(n-1)J_n(y)
\end{equation}
with $J_2=2\tan^{-1}\left(e^y\right)$, $J_3=\tanh y$ and where we have defined $J_n(y)=I_n(\sinh x)$.

 For non-retarded van der Waals forces,
 $n=6$, the potential $V(x,z;6)$ reproduces the expression (2) in Ref.~\cite{MP1}. For retarded van der Waals forces,
 $n=7$, on which we now focus, the integrals over $I_n(x)$ in (\ref{V2}) may also be done explicitly leading to
\begin{widetext}
\begin{equation}
V_2(x,z;7)=\frac{1}{15}\varepsilon_w\rho_w\sigma^7\frac{6\arctan\left(\frac{z}{x}\right)(x^6+x^4z^2-x^2z^4-z^6)-3\pi x^2z^4-3\pi
z^6-6x^5z-4x^3z^3-6xz^5}{x^4(x^2+z^2)z^4}\,,
\end{equation}
\end{widetext}
outside of the hard-wall domain. We stress that while the present mean-field treatment does not account for some of the fluctuation effects predicted for  wedge filling
-- in particular the roughness of the liquid-gas interface -- we expect that our DFT should be otherwise extremely accurate regarding the location of the transition, its
order and the divergence of the adsorption.

\begin{figure}
\includegraphics[width=0.45\textwidth]{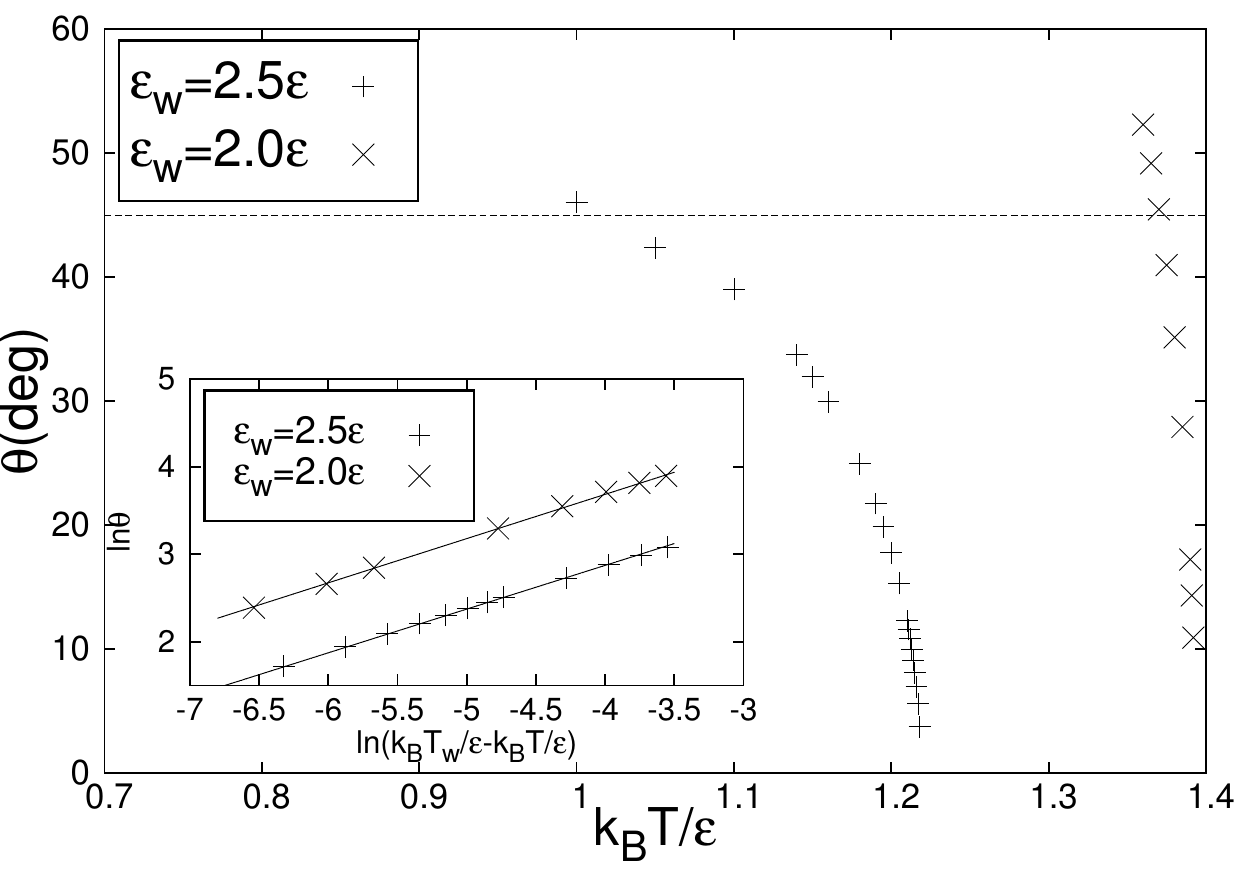}
\caption{Temperature dependence of the contact angle for two different wall strengths $\varepsilon_w=2\,\varepsilon$ and $\varepsilon_w=2.5\,\varepsilon$. The bulk critical temperature is located at $k_BT_c/\varepsilon=1.41$. The intersection of $\theta(T)$
with the dashed line at $\theta=45^\circ$ is the thermodynamic prediction for $T_{\rm fill}$. In the inset is shown a log-log plot illustrating the vanishing of $\theta$
in the vicinity of each $T_{\rm wet}$; both fitted straight lines have slope equal to $1/2$ consistent with the expected first-order singularity $\theta(T)\sim (T_{\rm
wet}-T)^{\frac{1}{2}}$.}\label{fig2}
\end{figure}

We discuss representative results for $\varepsilon_w=2.5\,\varepsilon$ and $\varepsilon_w=2\,\varepsilon.$ For each, we first considered the planar wall with potential
$V_\pi (z;7)$ and determined the contact angle $\theta(T)$ using Young's equation following the same method described in Ref.~\cite{wedge_long} (see Fig.~\ref{fig2}).
Each system exhibits a wetting transition with $T_{\rm wet}=0.86\,T_c$ and $T_{\rm wet}=0.98 \,T_c$ for the stronger and weaker wall respectively. Both wetting
transitions are first-order with the contact angle vanishing as $\theta(T)\approx (T_{\rm wet}-T)^{\frac{1}{2}}$ (see inset). This is expected since the wall-fluid
potential is long-ranged but the truncated LJ fluid-fluid interaction is short-ranged prohibiting continuous wetting \cite{dietrich}. In addition this must also mean
that the wetting binding potential $W(\ell)$ {\it{always}} exhibits an activation barrier even far below $T_{\rm wet}$; a change in sign of the Hamaker constant $A$ is
{\it{only}} possible if the wall-fluid and fluid-fluid forces have the same range. Thus there is no spinodal temperature $T_s$ at which the activation barrier
disappears. However for wetting (and filling) transitions occurring near $T_c$ the size of the activation barrier may be very small due to the small difference in bulk
liquid and gas densities. We will return to this shortly.

\begin{figure}
\includegraphics[width=0.4\textwidth]{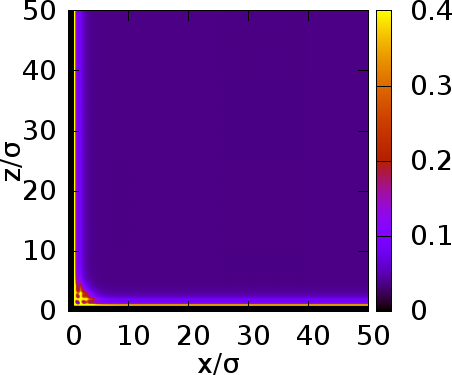}
\includegraphics[width=0.4\textwidth]{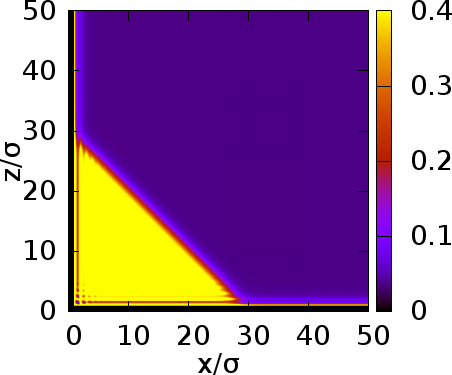}
\caption{Coexisting density profiles near a first-order filling transition for wall strength  $\varepsilon_w=2.5\,\varepsilon$ (corresponding
to $k_BT_{\rm fill}/\varepsilon=1.05$). The upper panel shows the microscopic configuration in which the interface is tightly bound to the apex. The lower panel shows
the macroscopic  configuration in which the meniscus is far from the wall and meets each wall at the contact angle $\theta=\pi/4$ in line with the macroscopic condition
(1). }\label{fig3}
\end{figure}

Turning attention to the right-angle wedge we determine equilibrium free-energies and density profiles at bulk coexistence from  minimization of the functional
$\Omega[\rho]$. This is done on an $L\times L$ grid (with $L=50\sigma$) with discretization size $0.05\,\sigma$ using the same numerical scheme described in
Ref.~\cite{wedge_long}. According to the thermodynamic prediction (\ref{thermo}), the location of the filling transitions can be determined from the intersection of the
contact angle curves with $\psi=\pi/4$ and gives $T_{\rm fill}=0.72\,T_c$ and $T_{\rm fill}=0.97\,T_c$ as $\varepsilon_w$ decreases in strength. These predictions are in
near perfect agreement with our numerical results obtained from minimization of $\Omega[\rho]$ which also determines the order of the filling transition. Starting from
different high density and low density states configurations near first-order filling converge to different equilibrium profiles. This is what is found for the stronger
wall, with the lower filling temperature, as illustrated in Fig.\ref{fig3} which shows coexisting density profiles corresponding to microscopic and macroscopic
adsorptions. Of course the size of this macroscopic state is limited by our numerical grid and scales with the system size $L$.

\begin{figure}
\includegraphics[angle=0, width={0.2\textwidth}]{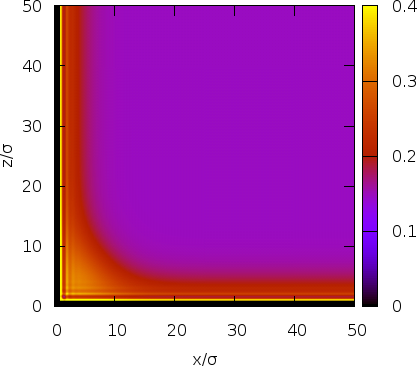}  \includegraphics[angle=0, width={0.2\textwidth}]{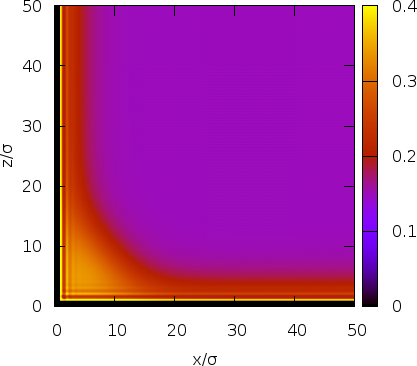}\\
\includegraphics[angle=0, width={0.2\textwidth}]{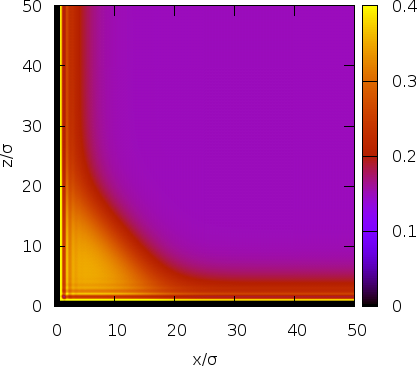} \includegraphics[angle=0, width={0.2\textwidth}]{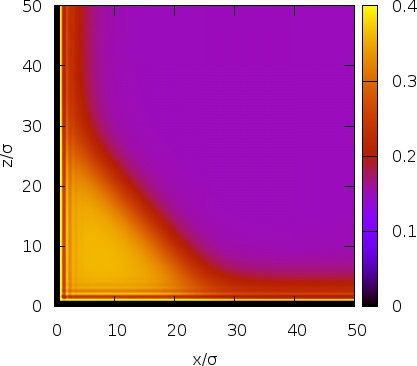}
\caption{Sequence of density profiles for the wedge with weaker strength ($\varepsilon_w=2\,\varepsilon$) showing a continuous increase in the adsorption of liquid as $T$ is increased to the filling transition occurring at $k_bT_{\rm fill}/\varepsilon=1.3693$. From top left to bottom right
$k_bT/\varepsilon=1.367, 1.368, 1.369$ and $1.3692$.
}\label{fig4}
\end{figure}

\begin{figure}
\includegraphics[width=0.5\textwidth]{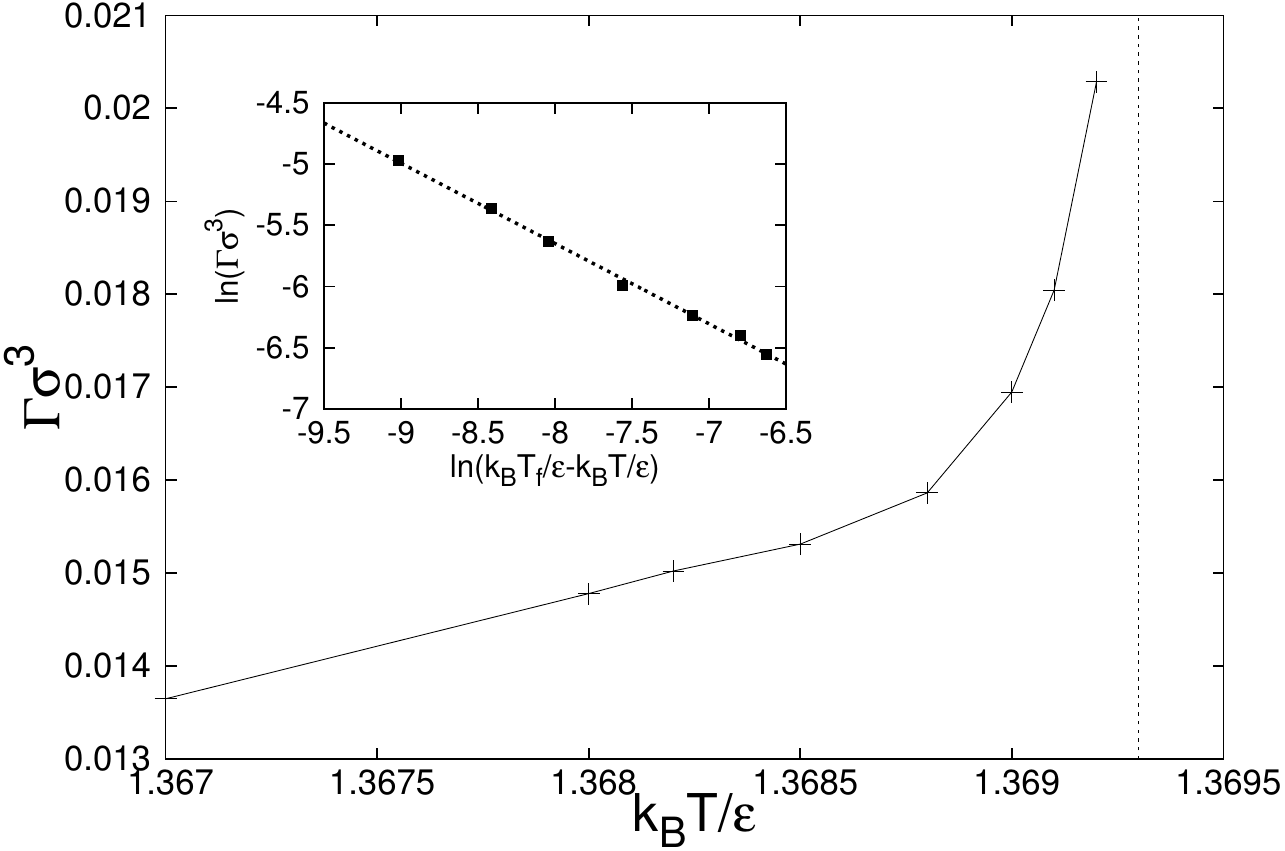}
\caption{Plot of the adsorption (in reduced units) as a function of temperature for the wedge with weaker wall strength $\varepsilon_w=2\,\varepsilon$. Inset is a log-log plot and comparison with the prediction for continuous filling with retarded van der Waals interactions $\Gamma\sim (T_{\rm fill}-T)^{-\frac{2}{3}}$. }\label{fig5}
\end{figure}

For the filling transition occurring at $T_{\rm fill}=0.97\,T_c$ on the other hand a unique phase is found for all temperatures indicating that the
transition is continuous. Corresponding density profiles are shown in Fig.\,\ref{fig4} and show the wedge gradually filling with liquid to a maximum
value determined by $L$. A plot of the adsorption $\Gamma$ versus $T$ is shown in Fig.\,\ref{fig5} and shows a dramatic but continuous increase in
the adsorption near the anticipated $T_{\rm fill}$. Thus, as with the earlier study with dispersion forces ($n=6$), it appears that within this
microscopic theory, the filling transition occurring near $T_c$ is continuous or at least effectively continuous. Assuming that the transition is
continuous the critical singularity for the adsorption is precisely in accord with the expectation (\ref{Gamma}). The inset in Fig.\,\ref{fig5} shows
a log-log plot of the adsorption for $T<T_{\rm fill}$, in which we use an {\it{unfitted}} estimate of the filling temperature obtained from
(\ref{thermo}) compared with  $\Gamma\sim (T_{\rm fill}-T)^{-\frac{2}{3}}$.
 For comparison, the same FM-DFT but with non-retarded forces ($n=6$) also showed critical point wedge filling yielding results
 consistent with the expected $p=2$ power-law $\Gamma\sim (T_{\rm fill}-T)^{-1}$ \cite{MP1,MP2}.

\begin{figure}
\includegraphics[width=0.5\textwidth]{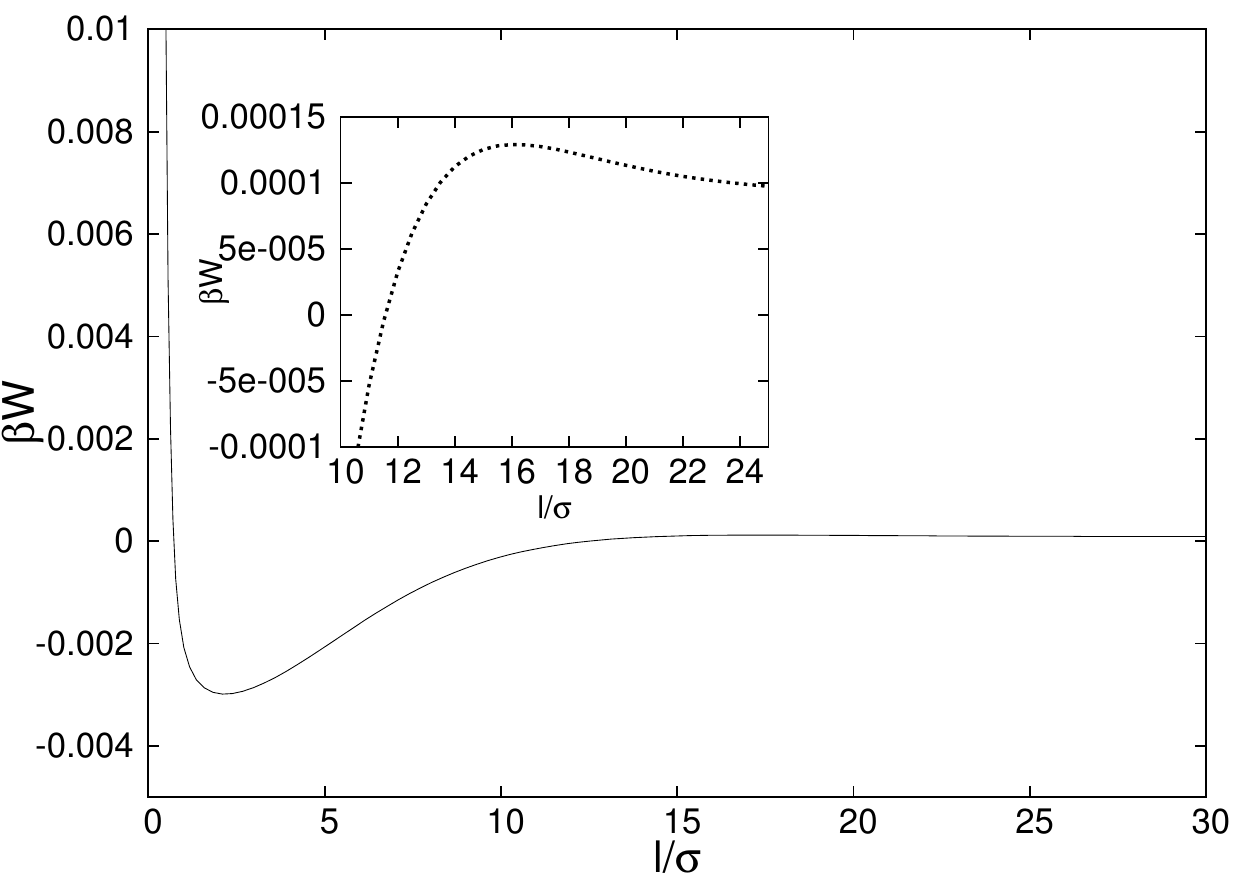}
\caption{Plot of the numerically determined binding potential for wetting at a planar wall with  $\varepsilon_w=2\varepsilon$ at $k_BT/\varepsilon=1.393$
close to the filling temperature $T_{\rm fill}$. Notice that $W(\ell)$ shows a very small activation barrier near $\ell\approx 16\sigma$.}\label{bind_pot}
\end{figure}

The observation of the expected $p$-dependence of the adsorption critical exponent for critical point wedge filling with retarded van der Waals
forces is the central result of our paper. This continuous phase transition occurs despite the fact that our model DFT has short-ranged fluid-fluid
interactions and that therefore the binding potential $W(\ell)$ for the planar wetting transition still exhibits an activation barrier. Therefore the
condition for continuous filling, according to the shallow wedge interfacial model is not met. However numerical determination of the binding
potential by constrained partial minimization of the Grand potential functional $\Omega[\rho]$ shows that the activation barrier is extremely small
near the filling temperature $T_{\rm fill}=0.97T_c$. This can be seen in our final figure (Fig.~\ref{bind_pot}) where the activation barrier is
hardly visible. Thus while the condition for continuous filling, according to the shallow wedge interfacial model, is not strictly met, the
transition would be predicted to be very weakly first-order. It is possible that in much larger systems the filling transition in our DFT model would
also appear first-order and that eventually the adsorption would show a jump to a macroscopic value. However this is {\it{not}} what was observed in
our previous studies for non-retarded forces and there is no indication of this in our numerical calculations. It appears to us to be more likely
that the shallow wedge interfacial model, which uses a purely local effective interfacial interaction is inadequate close to $T_c$, see
Ref.~\cite{MP2 for further discussion}. Whatever scenario, the observed behaviour in our numerical DFT study has all the hallmarks of a second-order
phase transition showing the expected $p$-dependence of the critical exponents. This means that the conditions for continuous or effective continuous
filling are even more relaxed than initially anticipated. This is something that we hope can be tested in the laboratory similar to experiments on
complete wedge filling \cite{mistura}. If continuous filling is experimentally accessible it would be a means of observing the dramatic and universal
enhancement of the interfacial roughness $\xi_\perp\propto (\theta-\alpha)^{-\frac{1}{4}}$ which is predicted even in the mean-field fluctuation
regime of wedge filling pertinent to systems with retarded or non-retarded forces \cite{wood2}.

\begin{acknowledgments}
 \noindent This work was funded in part by the EPSRC UK grant EP/L020564/1, ``Multiscale Analysis of Complex Interfacial Phenomena''.
 A.M. acknowledges the support from the Czech Science Foundation, project 16-12291S.
\end{acknowledgments}

\end{document}